\documentclass[pre,superscriptaddress,floatfix,longbibliography,twocolumn]{revtex4-1}

\usepackage{graphics}
\usepackage{graphicx}
\usepackage{amsmath}
\usepackage{amssymb,xcolor}
\usepackage{textcomp}
\usepackage{siunitx}
\usepackage{soul}

\newcommand\bv{\boldsymbol v}
\newcommand\bn{\boldsymbol n}

\newcommand{\Rb}{R_{b}}
\newcommand{\RD}{R_{D}}
\newcommand{\Rrim}{R_{\mathrm{rim}}}

\newcommand{\ez}{\hat{\bm{e}}_{z}}
\newcommand{\Dtens}{\bm{\mathcal{D}}}

\definecolor{grisclair}{rgb}{0.6,0.6,0.6}

\newcommand{\beq}{\begin{equation}}
\newcommand{\ee}{\end{equation}}

\begin{document}

\title{Bubble bursting in a sessile droplet}
\author{U. J. Guti\'errez-Hern\'andez}
\address{Departamento de F\'isica, Facultad de Ciencias, Universidad Nacional Aut\'onoma de México, Circuito Exterior S/N, Ciudad Universitaria, Ciudad de M\'exico 04510, M\'exico.}
\author{B. Muñoz-Sánchez}
\address{Depto.\ de Ingenier\'{\i}a Mec\'anica, Energ\'etica y de los Materiales and\\ 
Instituto de Computaci\'on Cient\'{\i}fica Avanzada (ICCAEx),\\
Universidad de Extremadura, E-06006 Badajoz, Spain}
\author{A. Agúndez-Cruz}
\address{Depto.\ de Ingenier\'{\i}a Mec\'anica, Energ\'etica y de los Materiales and\\ 
Instituto de Computaci\'on Cient\'{\i}fica Avanzada (ICCAEx),\\
Universidad de Extremadura, E-06006 Badajoz, Spain}
\author{J. M. Montanero}
\address{Depto.\ de Ingenier\'{\i}a Mec\'anica, Energ\'etica y de los Materiales and\\ 
Instituto de Computaci\'on Cient\'{\i}fica Avanzada (ICCAEx),\\
Universidad de Extremadura, E-06006 Badajoz, Spain}
\author{D. Fern\'andez Rivas}
\address{Mesoscale Chemical Systems Group, MESA+ Institute and Faculty of Science and Technology, University of Twente, PO Box 217, 7500 AE Enschede, The Netherlands}
\author{M. G. Cabezas}
\address{Depto.\ de Ingenier\'{\i}a Mec\'anica, Energ\'etica y de los Materiales and\\ 
Instituto de Computaci\'on Cient\'{\i}fica Avanzada (ICCAEx),\\
Universidad de Extremadura, E-06006 Badajoz, Spain}

\begin{abstract}
We analyzed experimentally and numerically the bursting of a bubble within a sessile droplet. Our experiments show that both sessile droplet curvature and confinement enhance the energy focusing. In the low-viscosity regime, this effect results in thinner, faster Worthington jets. In the high-viscosity regime, droplets are ejected for values of the Laplace number (the Reynolds number based on the visco-capillary velocity) smaller than the threshold for a bubble in an infinite liquid bath. This is probably the major result of the present work. Numerical simulations show the critical role of the additional pressure gradient arising from the curvature of the sessile droplet interface. The resulting force drives the liquid towards the bottom of the cavity, compressing it and accelerating jet formation. In the low-viscosity limit, the bottom of the cavity becomes smoother before jet ejection. This effect resembles the energy-focusing enhancement that occurs in an infinite liquid bath at the critical Laplace number, where short-wavelength waves are damped by viscosity. 
\end{abstract}

\maketitle

\section{Introduction}

Consider a bubble resting below the flat free surface of an infinite liquid bath. When the thin cap film covering the bubble ruptures, the collapse of the bubble generates capillary waves that converge at the cavity base, producing a Worthington jet whose tip ejects tiny droplets \citep{DPJZ02,GS16,G17a,BBWFYB18,DGLDZPS18,GR19,GL21}. This phenomenon has enormous consequences in diverse fields. Jet drops generated by the bursting of small bubbles at the ocean surface are widely recognized as a major source of aerosols with sizes around 0.1~$\mu$m \citep{V15b,D22,WDMR17}. These aerosols can act as cloud condensation nuclei in atmospheric regions where the water vapor concentration becomes supersaturated \citep{Cochran2017,Meshkhidze2019}. The drops can transport chemical pollutants, such as toxins and microplastics \citep{WDMR17,SLND23}, as well as biological agents including bacteria and viruses \citep{B21}, into the atmosphere, with important implications for public health \citep{Singh2026,Han2025}. The burst of an entrained bubble can produce the jumping of a centimeter-sized water droplet from a superhydrophobic surface \citep{HLYZCCSMCC26}. The impact of capillary waves in fluid-structure interactions can be leveraged for droplet actuation and directional particle printing in additive manufacturing \citep{HLYZCCSMCC26}.

Contaminants \citep{RCVMC26}, surfactants \citep{Baryiames2021,ND21,NED21,CKBSCJM21,PPS22,JYWEF23,VM24,YBJF24}, polymeric molecules \citep{SLJ21,TSKGMFM21,JYWEF23,RRMGC23,YBJF24,DOZLS25,BSJVT24,CMLCVM25,BYF25}, oils \citep{JYF21,YJF23,YJAF23,YLF25}, and particles of different nature \citep{JSF22,DMB23} have proved to substantially affect bubble bursting. These elements reduce the energy focusing responsible for the jet formation, thereby increasing the size of the first-emitted droplet and decreasing its velocity relative to their counterparts in a clean liquid \citep{DPJZ02,GS16,G17a,BBWFYB18,DGLDZPS18,GR19,GL21}. 

Almost concurrently with the work presented here, \citet{YSCF26} reported on the bursting of bubbles in shallow liquid layers. The nearby solid boundary reduces the size and increases the number of the emitted droplets due to a wall-induced viscous sticking effect that suppresses the upward motion of the cavity bottom. In this work, we propose another mechanism to enhance the focusing of energy on the Worthington jet ejected by the bursting bubble. The bubble is confined within a sessile droplet with the triplet contact line pinned to the surface. Our experiments show that trapping the released interfacial energy within the droplet decreases the size of the first-emitted droplet and increases its velocity in the inviscid regime. In the viscous regime, it enables liquid ejection for values of the Laplace number (the Reynolds number based on the visco-capillary velocity) below the minimum value in all previous experiments. Numerical simulations allow us to gain insight into the physical mechanisms responsible for these effects. 

The paper is organized as follows. We begin in Sec.\ \ref{sec2} with a brief summary of some results of bubble bursting in an infinite liquid bath. Our problem is formulated in dimensionless terms in Sec.\ \ref{sec3}. The experimental and numerical methods are described in Sec.\ \ref{sec4}. Section \ref{sec5} presents the experimental results, while the numerical simulations are presented in Sec.\ \ref{sec6}. Finally, the paper closes with some conclusions in Sec.\ \ref{sec7}.

\section{Bubble bursting in an infinite liquid bath}
\label{sec2}

Consider a bubble bursting at a flat interface far away from any solid surface. The liquid density and viscosity are $\rho$ and $\mu$, respectively, and the surface tension is $\sigma$. Neglecting the effects of air, the radius $R_d$ and velocity $v_d$ of the first droplet emitted by the jet after the bubble bursting obey the expressions  
\begin{eqnarray}
\frac{R_d}{\ell_{\mu}}=f_R(\text{Bo},\text{La}),\quad  \frac{v_d}{v_{\mu}}=f_v(\text{Bo},\text{La}),
\end{eqnarray}
where $\ell_{\mu}=\mu^2/(\rho\sigma)$ and $v_{\mu}=\sigma/\mu$ are the visco-capillary length and velocity, respectively, $\text{Bo}=\rho g R_b^2/\sigma$ is the Bond number, $\text{La}=\rho R_b \sigma/\mu^2$ is the Laplace number, and $R_b=[3V_b/(4\pi)]^{1/3}$ and $V_b$ are the bubble radius and volume, respectively. Finally, $g$ is the gravity.

\begin{figure}
\vspace{0.cm}
\begin{center}
\resizebox{0.37\textwidth}{!}{\includegraphics{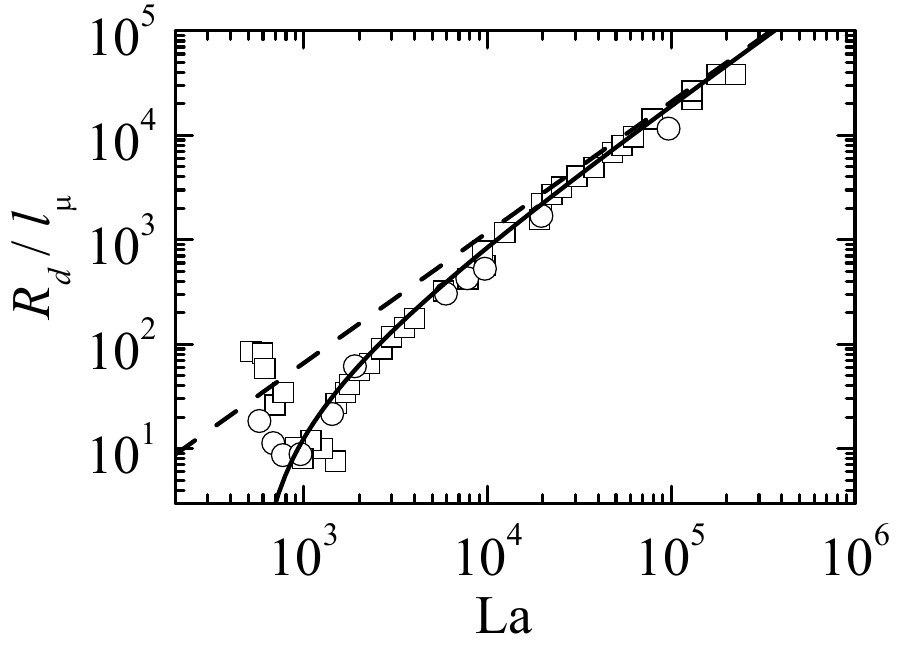}}
\end{center}
\caption{$R_d/\ell_{\mu}$ as a function of the Laplace number La. The solid line is the prediction obtained from the scaling law (\ref{law}). The dashed line corresponds to the high-Laplace number regime $\varepsilon\gg \text{La}^{-1/2}$. The symbols are simulation results for $\text{Bo}\leq 10^{-3}$ \citep{DGLDZPS18}.}
\label{scaling}
\end{figure}

For a fixed Bond number, the droplet radius decreases (the velocity increases) as the Laplace number decreases until reaching a minimum (maximum) (Fig.\ \ref{scaling}). The parameter conditions leading to the minimum size and maximum velocity correspond to $\text{Bo}=0$ and $\text{La}=\text{La}^*\simeq 1110$. The existence of an optimum value of the Laplace number, $\text{La}^*\simeq 1100$, has been attributed to an energy-focusing effect \citep{DGLDZPS18} during the bubble collapse: the mechanical energy per unit volume focused on the ejected ligament is maximized for that optimum value of the Laplace number \citep{GL21}. Short-wavelength capillary waves precede the capillary wave responsible for Worthington jet ejection. For $\text{La}>\text{La}^*$, those waves are damped by viscosity, enhancing the focusing of energy. For $\text{La}<\text{La}^*$, viscosity also dissipates the energy transported by the jet precursor wave, reducing the jet kinetic energy. Time-dependent profiles of the cavity collapse are self-similar at the critical Laplace number \citep{DPJZ02,LED18,GB23,GRS26}.  

Different scaling laws have been derived to describe bubble bursting in an infinite liquid bath \citep{G17a,GL21,LED18,DGLDZPS18,GR19,BDSP20}. Here, we use the scaling analysis proposed by \citet{G17a} as a relatively simple reference for comparing our results. According to this analysis, the droplet radius $R_d$ and velocity $v_d$ for a vanishing Bond number can be estimated as
\begin{equation}
\label{law}
\frac{R_d}{\ell_{\mu}}=k_d\, \phi^{5/4},\quad \frac{v_d}{v_{\mu}}=k_v\, \phi^{-3/4},    
\end{equation}
\begin{equation}
\label{law2}
\quad \phi=\left(\varepsilon-\text{La}^{-1/2}\right)\text{La},
\end{equation}
where $k_d=0.6$ and $k_v=16$ are determined by fitting (\ref{law}) to experimental and numerical results. The term $\varepsilon-\text{La}^{-1/2}$ indicates the excess of energy $\Delta E$ (in terms of the bubble interfacial energy $\sigma R_b^2$) available for the droplet detachment from the jet. In fact, $\Delta E=0$ allows one to estimate the minimum value of La for droplet ejection. The constant $\varepsilon=0.043$ indicates the value of $\Delta E$ without viscous dissipation, and the term $\text{La}^{-1/2}$ accounts for the portion of that energy dissipated by viscosity. This portion is estimated as $\Phi R_b^3 t_{cb}/(\sigma R_b^2)$, where $\Phi\sim \mu v_{cb}^2/R_b^2$ is the viscous dissipation function, $v_{cb}=R_b/t_{cb}$ is the characteristic (capillary wave) velocity, and $t_{cb}=(\rho R_b^3/\sigma)^{1/2}$ is the bubble bursting (inertio-capillary) time scale. The predictions (\ref{law}) agree with experimental and numerical results for $\text{La}>\text{La}^*$ (Fig.\ \ref{scaling}).

In the high-Laplace number regime, $\varepsilon\gg \text{La}^{-1/2}$, and the kinetic energy of the jet with length $R_b$ scales as
\begin{equation}
\label{jet}
\rho v_d^2 R_d^2\, R_b\sim \sigma R_b^2\varepsilon,
\end{equation}
where the jet radius and velocity are estimated as those of the first-emitted droplet [Eqs.\ (\ref{law}) and (\ref{law2})]. This implies that, as viscosity increases within this regime, the energy transferred to the jet remains constant, but the focusing effect is enhanced ($R_d$ decreases and $v_d$ increases) due to the low-energy short-wavelength capillary waves that precede jet ejection. 

Corrections to Eqs.\ (\ref{law}) due to gravity \citep{G18b,DGLDZPS18} are not relevant for our analysis due to the moderately low values of Bo in our experiments. \citet{GL21} extended their scaling analysis to cover the full range of the Laplace number. As mentioned above, other scaling laws have been proposed to successfully rationalize numerical and experimental results \citep{DGLDZPS18,GR19,BDSP20,M24}. 

\section{Formulation of the problem}
\label{sec3}

\begin{figure}
\vspace{0.cm}
\begin{center}
\resizebox{0.35\textwidth}{!}{\includegraphics{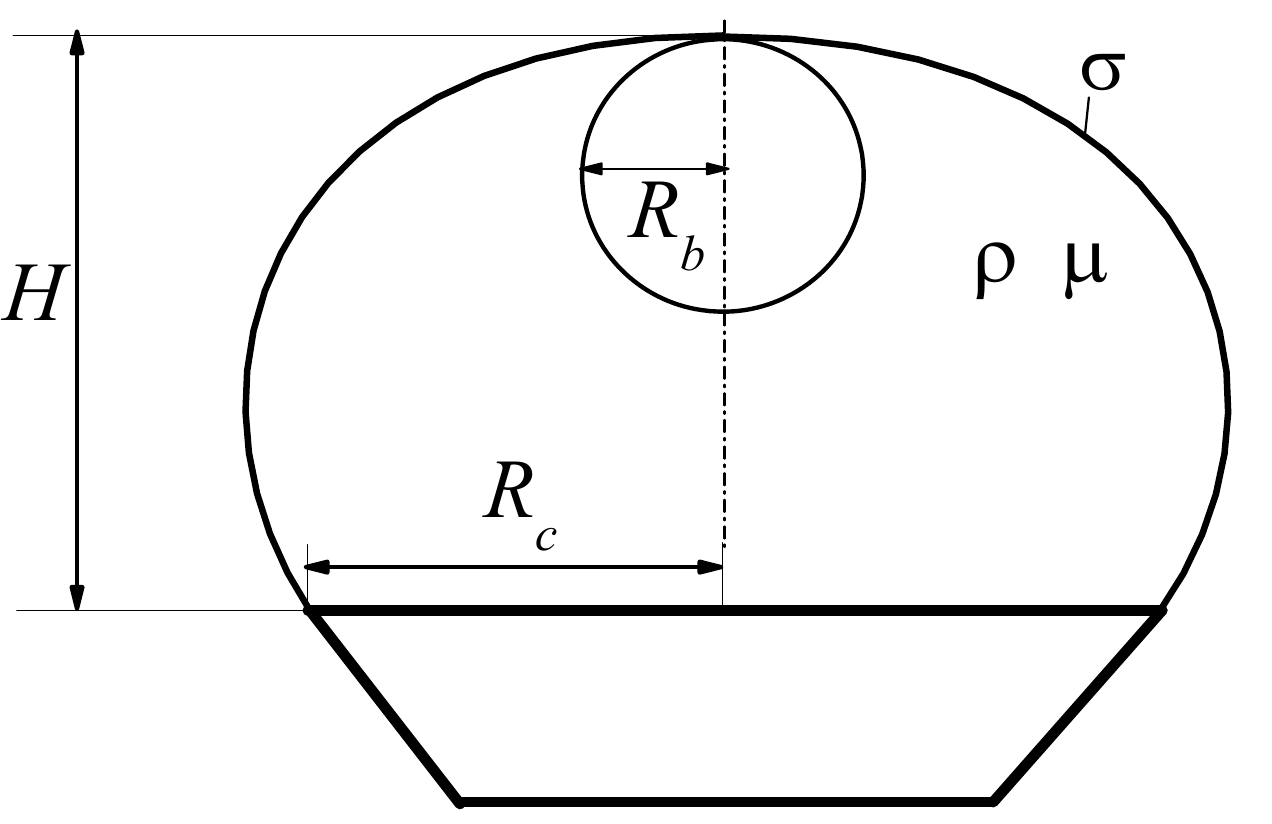}}
\end{center}
\caption{Sketch of the fluid configuration analyzed in this work. Parameters characterizing the bubble bursting in a sessile droplet.}
\label{sketch}
\end{figure}

Consider a sessile droplet of a liquid of density $\rho$ and viscosity $\mu$ resting on a solid surface (Fig.\ \ref{sketch}). The liquid-air surface tension is $\sigma$, the triple contact line radius is $R_D$, and the droplet volume is $V_i$. A bubble of volume $V_b$ is located at the droplet apex. The equivalent bubble radius is $R_b=[3/(4\pi) V_b]^{1/3}$. For a given liquid, the height $H$ of the droplet apex is a function of $R_D$, $V_i$, and $V_b$; i.e., $H=H(\rho,\sigma;R_D,V_i,V_b)$. We choose $R_D$, $H$, and $R_b$ as independent geometrical parameters. 

We assume that the triple contact line is pinned. Therefore, the wetting properties do not enter the problem. In our experiments, the triple contact line is anchored to the sharp edge of a solid disk. However, this is not necessary for the contact line to be pinned. The contact line tends to recede during bubble bursting on an infinite flat surface. In this case, the contact line pinning condition is equivalent to assuming a vanishing receding contact angle, characteristic of many rough surfaces \citep{Q08}.

We aim to determine the radius $R_d$ and velocity $v_d$ of the first droplet emitted by the jet after bubble bursting. Neglecting the effects of air, dimensional analysis dictates that  
\begin{eqnarray}
\frac{R_d}{\ell_{\mu}}&=&f_R(H/R_b,R_D/R_b;\text{Bo},\text{La}),\\
\frac{v_d}{v_{\mu}}&=&f_v(H/R_b,R_D/R_b;\text{Bo},\text{La}).
\end{eqnarray}
The classical problem of a bubble bursting at the surface of an infinite liquid bath corresponds to $H/R_b\to\infty$ and $R_D/R_b\to\infty$.

\section{Methods}
\label{sec4}

\subsection{Experimental method}

\begin{figure}
\vspace{0.cm}
\begin{center}
\resizebox{0.5\textwidth}{!}{\includegraphics{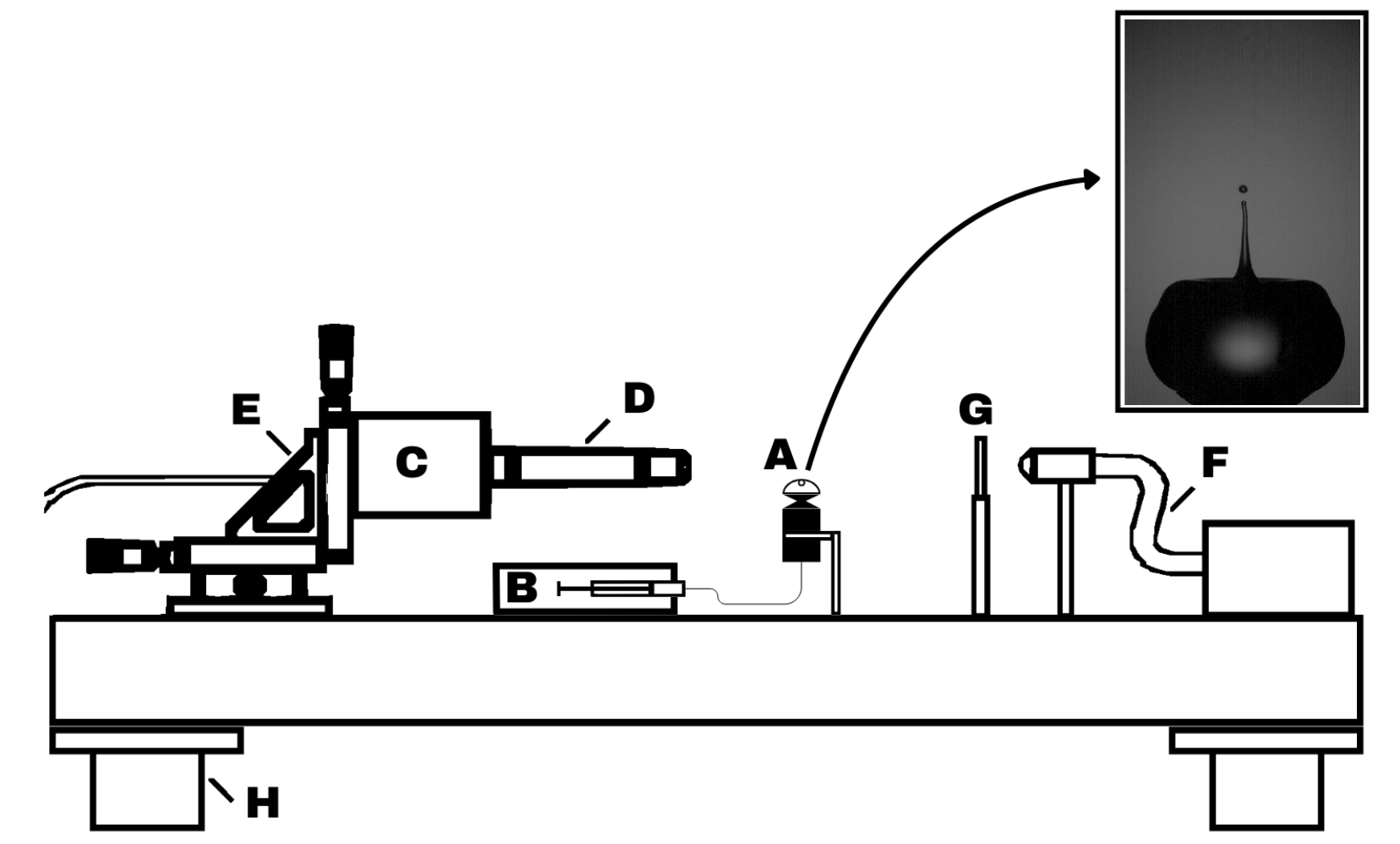}}
\end{center}
\caption{Experimental setup: pedestal (A), syringe pump (B), camera (C), optical lenses (D), triaxial translation stage (E), optical fiber (F), frosted diffuser (G), and pneumatic anti-vibration isolation system (H).}
\label{setup}
\end{figure}

The experimental setup consisted of a circular stainless-steel pedestal of radius $R_D=2$ mm. The pedestal had a sharp edge and an orifice of $\sim$ 100 $\mu$m in radius located in its center (Fig.\ \ref{setup}). Air can be injected through that orifice using a syringe pump (Harvard Apparatus PHD 4400). 

A high-speed video camera (either Fastcam Mini UX50 or Fastcam SA5 depending on the experiment) was used to capture bubble bursting. Images were recorded at a rate in the interval $2000-20000$ fps and with an exposure time in the interval $1.74-15.6$ $\mu$s depending on the experiment. The optical lenses provided a magnification in the interval $2.69-17.2$ $\mu$m/pixel. The number of pixels was selected according to the magnification. The camera could be displaced horizontally and vertically using two triaxial translation stages to focus the bubble. The bubble was illuminated from behind by cool white light emitted by two optical fibers. All these elements were mounted on an optical table equipped with a pneumatic anti-vibration isolation system to attenuate building-borne vibrations. Figure \ref{setup} shows the experimental setup.


Experiments were conducted with ultrapure water (W), a silicone oil with a viscosity of 5 cSt (SO), two aqueous solutions of sucrose at the concentrations 43.5\% (W/S/43.5) and 46.5\% (W/S/46.5) in weight, and octanol. This mixture enabled us to adjust the Laplace number to explore the viscous regime. It is worth noting that experiments with glycerol/water mixtures (commonly used in this type of experiment) may not be sufficiently reproducible \citep{RCVMC26}. The liquid properties are shown in Table \ref{properties}.

\begin{table}
    \centering
    \begin{tabular}{c|c|c|c}
         \begin{tabular}{c} Liquid \\          \end{tabular} &
         \begin{tabular}{c} Density \\  (kg/m$^3$)        \end{tabular} &   
         \begin{tabular}{c} Viscosity \\ (cSt)        \end{tabular} & 
         \begin{tabular}{c} Surface tension \\ (mN/m)        \end{tabular}
\\ \hline
W    & 998   & 1.00  &  69.3 \\
SO       &  913  &  5.00 &  19.7 \\
W/S/43.5 &	1201 &  6.20 &  74.2 \\
W/S/46.5 &	1218 &  8.51 &  74.9 \\
Octanol  &	827  &  8.71 &  23.5 \\
    \end{tabular}
    \caption{Properties of the liquids used in the experiments at 20 $^{\circ}$C.}
    \label{properties}
\end{table}

A sessile droplet was formed by depositing the working liquid onto the pedestal with a micropipette. The liquid expanded on the pedestal surface until the triple contact line reached the edge. An image of the droplet was acquired to determine its volume, $V_i$. In the water experiments, an air bubble was quasi-statically inflated within the droplet through the orifice at the center of the pedestal. This injection system produced bubbles with an approximately constant radius of $R_b\simeq 1.02\pm 0.05$ mm. In the experiments with silicone oil, aqueous sucrose solutions, and octanol, the bubble was injected laterally using a micropipette. The bubble radius varied over the interval $0.24\leq R_b\leq 1.03$ mm. We verified that liquid did not return to the pedestal orifice during the experiment.

\begin{figure}
\vspace{0.cm}
\begin{center}
\resizebox{0.35\textwidth}{!}{\includegraphics{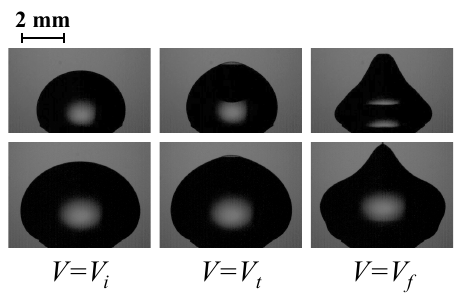}}
\end{center}
\caption{Images of the sessile droplet before injecting the bubble ($V=V_i$), after injecting the bubble ($V=V_t$), and after the liquid has been ejected ($V=V_f$). The experiments were conducted for for $R_D/R_b=1.96$, $\text{Bo}=0.15$, and $\text{La}=7.05\times10^{4}$. The upper and lower images correspond to $H/R_b=2.69$ and 3.86, respectively.}
\label{volumes}
\end{figure}

The bubble rose until it reached the droplet's apex, where it remained at rest for a few seconds. An image of the droplet was taken to determine the sum $V_t$ of the droplet volume $V_i$ and the bubble volume $V_b$. The bubble volume was calculated as $V_b=V_t-V_i$. Images of the Worthington jet were processed to determine the radius $R_d$ and velocity $v_d$ of the first emitted droplet. We also determined the droplet volume $V_f$ immediately after the jet emitted droplets. The volume emitted by the bubble bursting was calculated as $V_e=V_f-V_i$. The volumes were calculated by detecting the free surface at the sub-pixel resolution \citep{M24}. Figure \ref{volumes} shows the images used to calculate $V_i$, $V_t$, and $V_f$. All the liquid was ejected at the instant corresponding to the last image. The droplet's free surface oscillations are damped by viscosity until the droplet reaches equilibrium again. Determining the bubble radius is a significant source of uncertainty. This explains the considerable scattering of the experimental results. 

\subsection{Numerical method}
We complement the experiments with axisymmetric direct numerical simulations (DNS) of the post-rupture dynamics, performed with the open-source solver Basilisk~C \citep{popinet2009,popinet2015,basilisk}. The axisymmetric approximation is consistent with the experiment, in which the drop, the collapsing cavity and the resulting Worthington jet remain axisymmetric up to
the detachment of the first droplet. The liquid and the surrounding air are treated as incompressible Newtonian fluids in the one-fluid formulation \citep{prosperetti2009,tryggvason2011}. The interface is captured by a geometric, piecewise-linear (PLIC) Volume-of-Fluid (VoF) method, its curvature is evaluated with the height-function technique, and surface tension is applied through the balanced-force continuum surface force (CSF) scheme \citep{brackbill1992,popinet2009,popinet2018}. Momentum is advected consistently with the VoF-transported mass  \citep{popinet2018,basilisk}, which keeps the discretisation robust at the air--liquid density contrast. Gravity is implemented in its well-balanced (reduced-gravity) form, and the quadtree grid is adapted dynamically by wavelet-error estimation \citep{popinet2015,vanhooft2018}. Surface tension is taken to be uniform and constant.

All quantities are made dimensionless with the bubble radius $\Rb$, the inertio-capillary velocity $v_{cb}=(\sigma/\rho\Rb)^{1/2}$, the inertio-capillary time $t_{cb}=(\rho\Rb^{3}/\sigma)^{1/2}=\Rb/v_{cb}$ and the capillary pressure $\sigma/\Rb$ \citep{DPJZ02,DGLDZPS18,GR19,BSJVT24}. With these scales, and denoting by $f$ the liquid volume fraction, the dimensionless continuity and momentum equations read
\begin{gather}
\nabla\!\cdot\bv = 0 , \label{cont}\\[2pt]
\tilde{\rho}\!\left(\frac{\partial\bv}{\partial t}
  +\bv\!\cdot\!\nabla\bv\right)
= -\nabla p
  + \nabla\!\cdot\!\bigl(2\tilde{\mu}\,\Dtens\bigr)
  + \kappa\,\nabla f
  - \tilde{\rho}\,\text{Bo}\,\ez ,\label{mom}
\end{gather}
where $\Dtens=\tfrac{1}{2}[\nabla\bv+(\nabla\bv)^{\mathrm{T}}]$ is the strain-rate tensor, $\kappa$ is the interface curvature made dimensionless with $\Rb^{-1}$. The last term is the gravitational body force, with $\hat{\bm g}=-\text{Bo}\,\ez$, and $\ez$ the unit vector along the symmetry axis, pointing vertically upwards. Following the CSF model \citep{brackbill1992,popinet2009}, the singular capillary traction $\kappa\,\bn\,\delta_{s}$, in which $\bn$ is the unit normal to the interface and $\delta_{s}$ the interface Dirac distribution, is regularised as $\kappa\,\nabla f$. The dimensionless surface-tension coefficient is unity by construction of the scales. The mixture properties are $\tilde{\rho}=f+(1-f)\rho_{r}$ and $\tilde{\mu}=\text{La}^{-1/2}\,[\,f+(1-f)\mu_{r}\,]$, so that the dimensionless liquid viscosity is the Ohnesorge number $\text{Oh}=\text{La}^{-1/2}=\mu/(\rho\sigma\Rb)^{1/2}$. The DNS retains the gas phase, with $\rho_{r}=\rho_{g}/\rho\sim10^{-3}$ and $\mu_{r}=\mu_{g}/\mu\sim10^{-2}$ for the liquids used. The flow is therefore governed by the Bond and Laplace numbers and, for the sessile drop, by the confinement parameters $H/\Rb$ and $\RD/\Rb$ introduced in \S\,\ref{sec3}. Being small, $\rho_{r}$ and $\mu_{r}$ have a negligible influence on the liquid dynamics, consistent with the neglect of air in the scaling analysis of \S\,\ref{sec2}. For every case, $\text{Bo}$ and $\text{La}$ are evaluated from the density, viscosity and surface tension measured for each liquid (Table~\ref{properties}), so that each simulation corresponds one-to-one to an experiment.

The simulations start an instant after the rupture of the bubble cap, the thin film is removed at $t=0$ and the flow begins from rest, as is standard for bubble bursting \citep{DPJZ02,DGLDZPS18,GR19,BSJVT24}. For the sessile drop, the initial shape is obtained by solving the coupled Young--Laplace equilibrium of the outer meniscus of the drop and of the bubble wall, joined at the edge of the spherical-cap film, with the contact line pinned at $r=\RD$ and the liquid and gas volumes $V_{i}$ and $V_{b}$ prescribed, the apex height $H$ then follows from the equilibrium. 
For the infinite bath, the classical equilibrium of a bubble at a flat free surface \citep{lhuissier2012,DGLDZPS18} is used; it is the $H/\Rb\to\infty$, $\RD/\Rb\to\infty$ limit of the sessile configuration. In both geometries, the edge left by the removed film is regularised by a small torus of radius $\Rrim$, tangent to the two interface arcs it joins, which represents the liquid collected by the Taylor--Culick retraction of the film \citep{taylor1959,culick1960}. It is important to note that $\Rrim$ is a regularisation parameter, it stands in for the film thickness and for the instant at which the numerical clock is started, neither of which is resolved here. We therefore choose $\Rrim/\Rb$ in the range $0.01$--$0.03$, which is small enough for the rim volume to be a negligible fraction of the bubble volume, large enough for the rim to be described by $7$--$20$ cells at the working resolution, and consistent with the Taylor--Culick collection of a film of thickness \citep{lhuissier2012}. The same $\Rrim$ is used for the bath and for the sessile drop at matched ($\text{Bo}$, $\text{La}$). 

The equations are solved in a square domain of side $12\,\Rb$. For the sessile drop, the lower boundary is an impermeable no-slip wall representing the pedestal, on which the contact line is pinned by a zonal condition on the volume fraction that imposes $f=1$ for $r<\RD$ and $f=0$ for $r>\RD$. This prescribes the position of the contact line rather than the contact angle, i.e.,  the liquid footprint is clamped at $r=\RD$ without invoking any static or dynamic contact angle model, while the apparent contact angle remains free to adjust to the local dynamics. This mimics the experimental contact line anchored at the sharp edge of the pedestal. The upper and lateral boundaries are outflow boundaries. For the bath, the far-field boundaries below and beside the cavity are free-slip walls \citep{BSJVT24} placed such that the clearance from the cavity exceeds $4\,\Rb$ at all analysed times, so the infinite-bath regime is realised. Apart from the initial condition and these boundary conditions, the two configurations are integrated with the same solver and the same refinement criteria. 

The grid is adapted at every time step by wavelet-error estimation \citep{popinet2015,vanhooft2018} on $\{f,v_{z},v_{r},\kappa\}$ with tolerances $\{10^{-3},10^{-2},10^{-2},10^{-8}\}$. The tight curvature tolerance drives the mesh to the maximum level along the entire interface. With a maximum level of $13$ over the $12\,\Rb$ box, the minimum cell size is $\Delta_{\min}=12\,\Rb/2^{13}\simeq\Rb/683$, which is consistent with recent bubble-bursting DNS works \citep{HLYZCCSMCC26,BSJVT24,BSJVT24}. The time step is limited by a CFL condition and by the explicit capillary stability constraint $\Delta t<(\bar{\rho}\,\Delta_{\min}^{3}/\pi)^{1/2}$, where $\bar{\rho}$ is the mean of the two phase densities \citep{brackbill1992,popinet2009}, and the Poisson--Helmholtz problems are converged to a relative tolerance of $10^{-4}$.  Grid convergence was assessed by doubling the resolution to $\Delta_{\min}\simeq\Rb/1365$ obtaining a change of less than $2\%$ in the radius and the speed of the first ejected droplet. 

Detached droplets are identified at each output time by a connected-component algorithm applied to the region $f>1/2$. The radius of the first ejected droplet is obtained from its volume as $R_{d}=[3V_{d}/(4\pi)]^{1/3}$ and its velocity $v_{d}$ from the axial velocity of its centre of mass, both evaluated immediately after detachment. Because the pinch-off of an axisymmetric ligament is ultimately regularised by the mesh, these are the two quantities on which the grid-convergence test above was performed. Simulations of the infinite bath reproduce both the droplet size and velocity predicted by \eqref{law} and the threshold Laplace number below which no droplet is ejected (figure~\ref{scaling}). The kinetic, interfacial, gravitational and cumulative viscous-dissipation energies discussed in the text are computed as detailed in the Supplemental Material, where the residual of the energy budget is also reported.

\section{Experimental results}
\label{sec5}

\subsection {High-Laplace number regime}

First, we consider experiments with water, in which $\text{La}\sim 10^5$. They correspond to the high-Laplace number regime $\varepsilon\gg \text{La}^{-1/2}$ of the infinite liquid bath case, where viscous dissipation can be neglected. 

\begin{figure*}
\resizebox{0.7\textwidth}{!}{\includegraphics{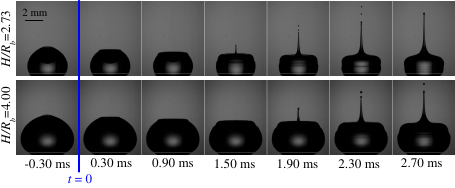}}
\caption{Images of the bubble bursting for $R_D/R_b=1.96$, $\text{Bo}=0.15$, and $\text{La}=7.05\times10^{4}$. The origin of time corresponds to the breakage of the cap film.}
\label{tiras}
\end{figure*}

Figure \ref{tiras} shows images of the bubble bursting and the liquid ejection for two representative values of the confinement parameter $H/R_b$. The jet becomes thinner and faster as $H/R_b$ decreases. Most of the droplet surface remained unperturbed when the jet appeared above the drop surface. This can be anticipated from the difference between the capillary time $t_{cb}=(\rho R_b^3/\sigma)^{1/2}=3.8$ ms characterizing the cavity collapse, and that of the sessile droplet $t_{cD}=(\rho R_D^3/\sigma)^{1/2}=10.5$ ms. The volume of the first-emitted droplet was smaller for $H/R_b=2.69$ by 43\%, while the velocity was larger by 41\%.

\begin{figure}
\resizebox{0.36\textwidth}{!}{\includegraphics{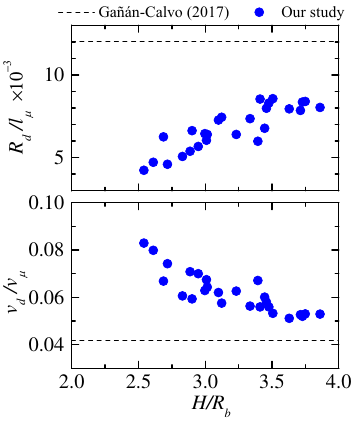}}
\caption{$R_d/\ell_{\mu}$ and $v_d/v_{\mu}$ as a function of $H/R_b$ for $R_D/R_b=1.96$, $\text{Bo}=0.15$, and $\text{La}=7.05\times10^{4}$. The dashed line is the prediction obtained from the scaling law (\ref{law}) for an infinite liquid bath.}
\label{water}
\end{figure}

As mentioned in Sec.\ \ref{sec3}, the air bubbles injected in the water droplets had approximately the same radius $R_b=1.02$ mm, implying that these experiments were conducted for essentially the same values of $R_D/R_b$, $\text{Bo}$, and $\text{La}$. This allows us to explore the dependence of $R_d/\ell_{\mu}$ and and $v_d/\ell_{\mu}$ on $H/R_b$. Figure \ref{water} shows that $R_d$ decreases and $v_d$ increases as $H/R_b$ decreases, implying that confinement enhances energy focusing. As shown in Sec.\ \ref{sec6}, this effect can be attributed to the sessile droplet's capillary pressure, i.e., the increase in liquid pressure due to the droplet's curved surface. For the smallest value of $H/R_b$, the first-emitted droplet radius decreases by more than a factor of 3 with respect to the infinite liquid bath case, and the velocity increases by more than a factor of 2. The radius and velocity of the first-emitted droplet were in the ranges $61-124$ $\mu$m and $3.53-5.76$ m/s, respectively.

\begin{figure}
\vspace{0.cm}
\begin{center}
\resizebox{0.38\textwidth}{!}{\includegraphics{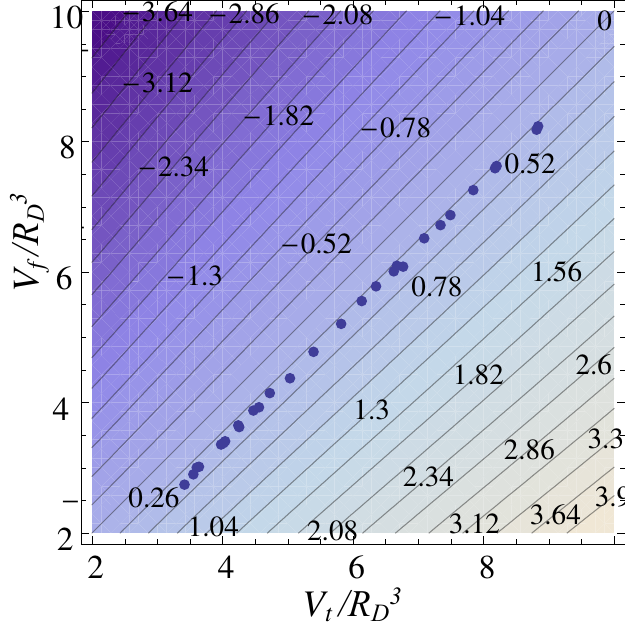}}
\end{center}
\caption{Isolines of $E_{st}/(4\pi R_b^2\, \sigma)$ as a function of the droplet volumes $V_t$ and $V_f$ before and after the bursting of a bubble for $\text{Bo}_d=\rho R_D^2 g/\sigma=0.544$. The symbols correspond to the experimental realizations.}
\label{edroplet}
\end{figure}

Bubble bursting in a sessile droplet entails the breakup of the droplet's outer surface. Figure \ref{edroplet} shows the total interfacial energy $E_{st}$ released by the bursting of the bubbles in water (the Supplemental Material describes the calculation of $E_{st}$). As can be observed, this energy is significant in terms of the bubble interfacial energy $4\pi R_b^2\, \sigma$ ($0.26\leq E_{st}/(4\pi R_b^2\, \sigma)\leq 0.52)$. Part of $E_{st}$ is transferred to the jet, potentially affecting the first-emitted droplet size and velocity. However, $E_{st}$ is delivered on the droplet time scale $t_{cD}=(\rho R_D^3/\sigma)^{1/2}$. Therefore, only a fraction of $E_{st}$ is transferred to the jet over the bubble time scale $t_{cb}=(\rho R_b^3/\sigma)^{1/2}=t_{cD} (R_b/R_D)^{3/2}$ ($t_{cb}<t_{cD}$). The rest is dissipated by viscosity on the much longer viscous time scale $t_{\mu D}=\rho R_D^2/\mu$. For a low viscosity liquid ($\text{Oh}_D=\mu/(\rho R_D\sigma)^{1/2}\ll 1)$, $t_{cD}\ll t_{\mu D}$. We will return to this point in Sec.\ \ref{sec6}.

\begin{figure}
\begin{center}
\resizebox{0.36\textwidth}{!}{\includegraphics{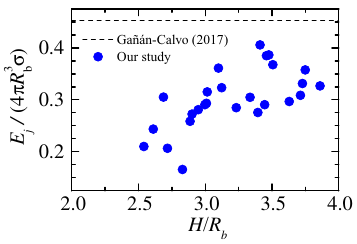}}
\end{center}
\caption{$E_j/(4\pi R_b^2 \sigma)$ as a function of $H/R_b$ for $R_D/R_b=1.96$, $\text{Bo}=0.15$, and $\text{La}=7.05\times10^{4}$.}
\label{water2}
\end{figure}

\begin{figure}
\resizebox{0.36\textwidth}{!}{\includegraphics{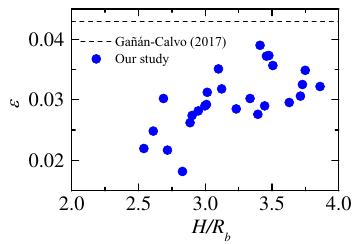}}
\caption{$\varepsilon$ as a function of $H/R_b$ for $R_D/R_b=1.96$, $\text{Bo}=0.15$, and $\text{La}=7.05\times10^{4}$. The dashed line indicates the value $\varepsilon=0.043$ for an infinite liquid bath \citep{G17a}.}
\label{epsilon}
\end{figure}

Confinement traps part of the energy $E_{st}$ released by the droplet surface after bubble bursting, thereby increasing the energy transferred to the jet. However, there are also two ``negative"\ effects: (i) the droplet surface oscillations disrupt energy transfer to the jet, and (ii) confinement increases viscous dissipation due to both triple-contact-line pinning and the solid surface. A natural question is how the energy transferred to the jet compares with that in the infinite liquid bath case. Figure \ref{water2} shows the dependence of the energy $E_j=1/2\, \rho\, \pi R_d^2 R_b v_d^2$ of a jet of length $R_b$ on the confinement parameter $H/R_b$ (as mentioned in Sec.\ \ref{sec2}, $H/R_b\to \infty$ corresponds to an infinite liquid bath). Here, the jet radius and velocity are estimated as those of the first-emitted droplet (Fig.\ \ref{water}), as in Eq.\ (\ref{jet}). As can be observed, the portion $E_j/(4\pi R_b^2 \sigma)$ of bubble interfacial energy $4\pi R_b^2 \sigma$ transferred to the jet decreases as $H/R_b$ decreases (confinement increases). This shows the ``negative" confinement effects mentioned above (energy transfer disruption and increased viscous dissipation). The value of $\varepsilon$ calculated from Eqs.\ (\ref{law}) and (\ref{law2}) with $R_d$ and $v_d$ obtained from our experiment are consistent with the limit $\varepsilon\to 0.043$ for $H/R_b\to \infty$ (Fig.\ \ref{epsilon}).

\begin{figure}
\resizebox{0.38\textwidth}{!}{\includegraphics{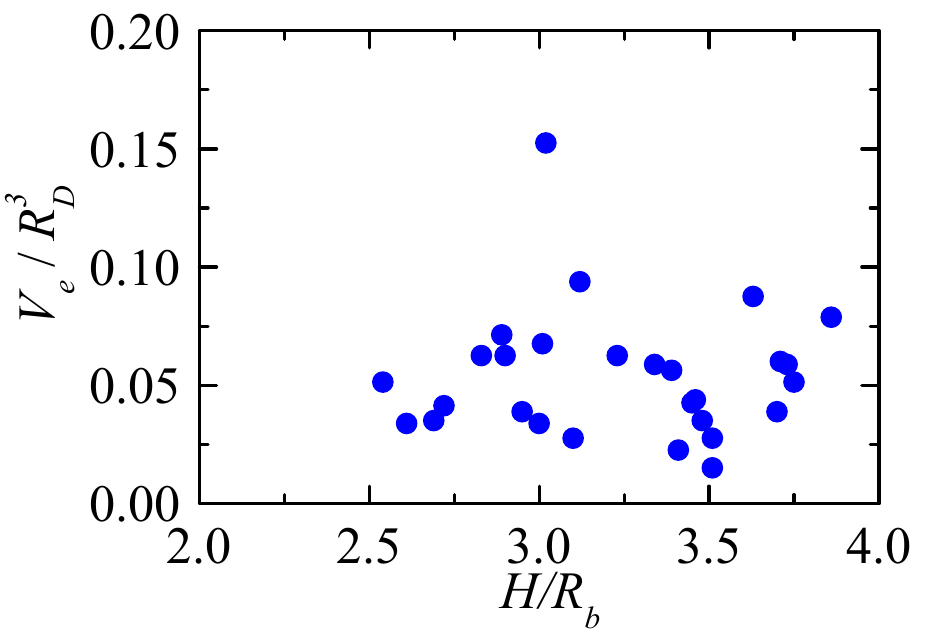}}
\caption{Total emitted volume $V_e/R_D^3$ as a function of $R_D/R_b=1.96$, $\text{Bo}=0.15$, and $\text{La}=7.05\times10^{4}$.}
\label{vole}
\end{figure}

As occurs in an infinite liquid bath, the total liquid volume emitted by the bubble bursting, $V_e=V_i-V_f$, is much smaller than the bubble volume ($V_e/R_D^3\lesssim 0.1$ and $V_b/R_D^3\simeq 0.57$), and does not exhibit a significant dependence on $H/R_b$ within the range analyzed in our experiments (Fig.\ \ref{vole}). The first-droplet volume averaged over all the experiments is $\langle V_d\rangle/R_D^3=5\times 10^{-4}$, only about 1\% of the average total liquid volume $\langle V_e\rangle/R_D^3\simeq 0.054$ emitted by the bubble bursting. We conclude that $V_d\ll V_e\ll V_b$, as also occurs in the infinite bath case.

\begin{figure}
\resizebox{0.36\textwidth}{!}{\includegraphics{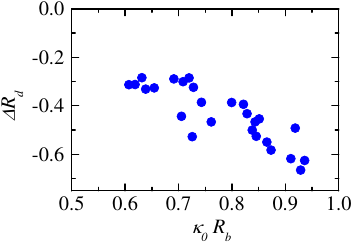}}
\caption{$\Delta R_d$ as a function of $\kappa_0 R_b$ for $R_D/R_b=1.96$, $\text{Bo}=0.15$, and $\text{La}=7.05\times10^{4}$.}
\label{delta}
\end{figure}

The curvature of a sessile droplet increases the pressure of the liquid surrounding the bubble relative to that in the infinite-bath case. We hypothesize that this increase is responsible for the reduction of the jet diameter and, therefore, the first-emitted droplet radius. To verify this, we calculated the decrease $\Delta R_d=(R_d-R_{d0})/R_{d0}$ in the first-emitted droplet radius $R_d$ with respect to the corresponding value $R_{d0}$ for zero free surface curvature, calculated from the scaling law (\ref{law}). Figure \ref{delta} shows the results as a function of the droplet apex curvature prior to bubble injection, $\kappa_0$, measured in terms of that of the bubble, $R_b^{-1}$. Notice that $\kappa_0 R_b$ is the capillary pressure at the droplet apex, $\sigma\kappa_0$, in terms of the driving capillary pressure $\sigma/R_b$ associated with the bubble interface. As shown in Fig.\ \ref{delta}, both quantities are commensurate with each other. In addition, $\Delta R_d$ is commensurate with $\kappa_0 R_b$ and exhibits an approximately linear dependence on that quantity.

To summarize, the results presented in this section show that energy focusing on the ejection of a low-viscosity liquid is enhanced in a sessile droplet due to the capillary pressure associated with the curved droplet surface. The first droplet emitted by the sessile droplet is significantly faster than that ejected by the infinite bath, despite the increase in viscous dissipation. This mechanism differs from that of a bubble bursting on a flat interface near a solid wall \citep{YSCF26}. In their case, the presence of the solid wall modifies the local velocity field and introduces viscous effects that hinder cavity shrinkage, leading to a steeper cavity geometry at the onset of jet formation.

\subsection {Low-Laplace number regime}

\begin{figure}
\resizebox{0.4\textwidth}{!}{\includegraphics{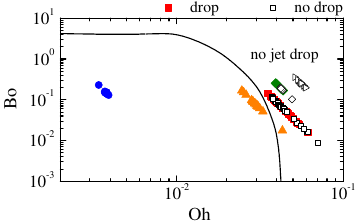}}
\caption{Regime map in the parameter plane defined by the Bond number Bo and Ohnesorge number $\text{Oh}=\text{La}^{-1/2}$. The solid line is the boundary of the droplet emission region for bubbles bursting in an infinite liquid bath \citep{WHB15}. The symbols correspond to experiments of a bubble inside a sessile drop of water ($\circ$), 5 cSt silicone oil ($\diamond$), W/S/43.5 ($\triangle$), W/S/46.5 ($\square$), and octanol ($\triangleright$). The solid (open) symbols correspond to experiments with (without) jet droplet emission.}
\label{BOh_map}
\end{figure}

Now we focus on the experiments conducted with viscous liquids. The solid line in Fig.\ \ref{BOh_map} delimits the droplet emission region for a bubble bursting in an infinite bath. Following \citet{WHB15}, we consider the Ohnesorge number $\text{Oh}=\text{La}^{-1/2}$ instead of La. Gravity and viscosity help prevent liquid from being emitted \citep{WHB15}. The effect of viscosity on the emission limit becomes significant for $\text{Oh}>10^{-2}$ ($\text{La}<10^{4}$), and no emission occurs for $\text{Oh}\gtrsim 0.043$ ($\text{La}\lesssim 540$). 

Bubble bursting in a sessile droplet produces jet drops within the emission region for an infinite bath (Fig.\ \ref{BOh_map}). Interestingly, the liquid confinement also allows for the production of droplets outside that region. In particular, liquid is emitted even for $\text{La}<540$. This is probably the major result of the present work. Notice that Fig.\ \ref{BOh_map} is a 2D projection of the 4D parameter space defined by the set of parameters $\{H/R_b,R_D/R_b;\text{Bo},\text{La}\}$. This explains why the droplet may or may not eject liquid for given values of Bo and La. 

\begin{figure}
\resizebox{0.4\textwidth}{!}{\includegraphics{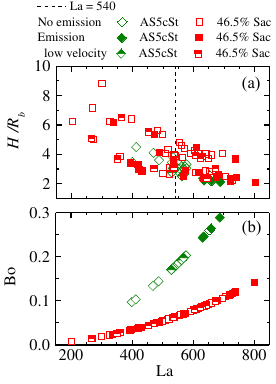}}
\caption{$H/R_b$ and Bo versus La for the experiments with viscous liquids. Experimental realizations with (solid symbols) and without (open symbols) emission. The half-filled symbols indicate that emission did not occur in all the experimental realizations.}
\label{Fig_EmisionLimit}
\end{figure}

Figure \ref{Fig_EmisionLimit} shows additional projections of the experimental data from experiments with viscous liquids. The general trend observed in Fig.\ \ref{Fig_EmisionLimit}a is that increasing confinement (decreasing $H/R_b$) increases the range of Laplace number for which emission occurs. In other words, emission occurs when $R_b/H$ (confinement) exceeds a threshold for a given value of the Laplace number. The phenomenon exhibits greater variability than its infinite-bath counterpart, owing to the higher dimensionality of the parameter space and the strong influence of the bubble radius, whose experimental uncertainty becomes significant for small bubbles (low Laplace numbers). Gravity collaborates with viscosity to suppress liquid ejection \citep{WHB15}. For this reason, decreasing the Bond number increases the range of the Laplace number for which ejection occurs (Fig.\ \ref{Fig_EmisionLimit}b).

\begin{figure*}
\resizebox{0.7\textwidth}{!}{\includegraphics{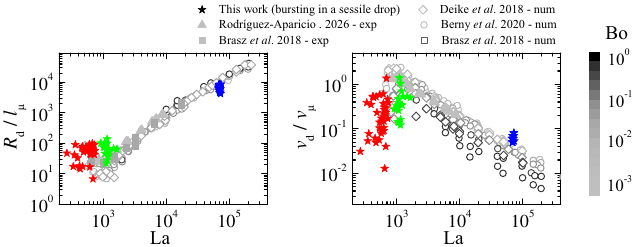}} 
\caption{Radius $R_d$ and velocity $v_d$ of the first-emitted droplet produced by bubble bursting in a sessile bubble (stars) and in an infinite bath (grey symbols).}
\label{Fig_AllResults}
\end{figure*}

In a infinite liquid bath, the smallest {\em experimental} value of the ratio $R_d/\ell_{\mu}$ is slightly above 20, which corresponds to $\text{Bo}\approx 0.1$ and $\text{La}=\text{La}^*\simeq 1110$ \citep{RCVMC26}. Smaller droplet radii were obtained in our experiments. Specifically, experiments with 5 cSt silicone oil emitted droplets with $R_d/\ell_{\mu}\simeq 7$  (Fig.\ \ref{Fig_AllResults}). This can be attributed to the enhanced energy focusing due to liquid confinement. 

Figure \ref{Fig_AllResults} shows that confinement smooths the minimum (maximum) of the curve $R_d/\ell_{\mu}(\text{La})$ ($v_d/\ell_{\mu}(\text{La})$) at the critical Laplace number $\text{La}=\text{La}^*\simeq 1100$ for an infinite bath. In an infinite bath, short-wavelength capillary waves preceding the one responsible for Worthington jet ejection are damped by viscosity for $\text{La}>\text{La}^*$, enhancing the focusing of energy. For $\text{La}<\text{La}^*$, viscosity also dissipates the energy transported by the jet precursor wave, reducing the jet kinetic energy. The minimum of $R_d/\ell_{\mu}(\text{La})$ results from the competition between these two opposite effects. The additional viscous dissipation in the solid surface and the pinned contact line of the sessile droplet may be responsible for smoothing the minimum of $R_d/\ell_{\mu}(\text{La})$. On the other hand, increased capillary pressure and confinement of the released interfacial energy within the droplet enable droplet ejection below the minimum Laplace number for an infinite bath. The Supplemental Material shows additional experimental results.

It is worth noting that viscoelasticity can affect the emission of smaller droplets of 5 cSt silicone oil and the aqueous sucrose solutions. The inertio-capillary time $t_{ic}=(\rho R_d^3/\sigma)^{1/2}$ characterizing the interface reversal yielding the droplet emission becomes of the order of a few microseconds. It has been shown that 5 cSt silicone oil may exhibit viscoelastic behavior on that timescale \citep{RPVHM19}. 

\begin{figure*}
\resizebox{0.7\textwidth}{!}{\includegraphics{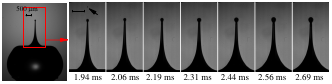}}
\caption{Worthington jet emitted by bubble bursting in a droplet of the W/S/46.5 for $H/R_b=4.47$, $R_D/R_b=2.79$, $\text{Bo}=0.082$, and $\text{La}=610$.}
\label{Fig_LowDrop10}
\end{figure*}

Figure \ref{Fig_LowDrop10} shows that the emission of the droplet of aqueous sucrose solution is followed by the growth of a thicker jet, which ``freezes" and no longer ejects liquid. This effect is probably due to extensional thickening induced by viscoelasticity. A similar behavior was found by \citet{CMLCVM25} when a bubble burst on the free surface of an infinite polymer solution bath.

\section{Numerical results}
\label{sec6}

\begin{figure*}
\resizebox{0.7\textwidth}{!}{\includegraphics{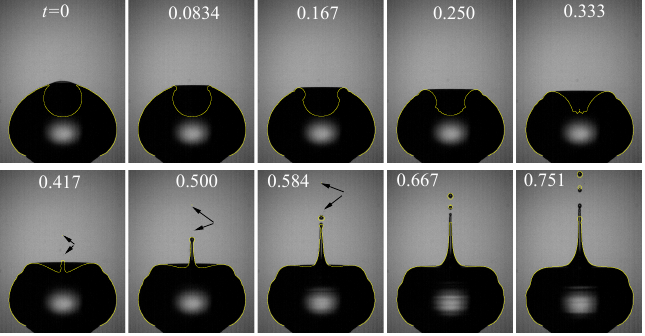}}
\caption{Comparison between the interface shape calculated numerically (solid lines) and obtained experimentally (images) for $H/R_b=3.94$, $R_D/R_b=1.99$, $\text{Bo}=0.143$, and $\text{La}=6.94\times10^{4}$. The arrows point to the ejected droplets.}
\label{Fig_SimVsExp}
\end{figure*}

We conducted numerical simulations to gain insight into the physical mechanisms responsible for the confinement effects described in the experimental section. Hereafter, the time, spatial coordinates, and hydrostatic pressure are expressed in terms of the inertio-capillary time $t_{cb}$, the bubble radius $R_b$, and the capillary pressure $\sigma/R_b$, respectively. Numerical simulations remarkably agree with the experiments, as illustrated in Fig.\ \ref{Fig_SimVsExp}. Interestingly, the tiny droplet emitted at $t\simeq 0.417$ was observed in both the experiment and the numerical simulation. This simulation shows that the droplet comes from the free-surface reversal produced by a capillary wave preceding the main one.

\begin{figure*}
\resizebox{0.8\textwidth}{!}{\includegraphics{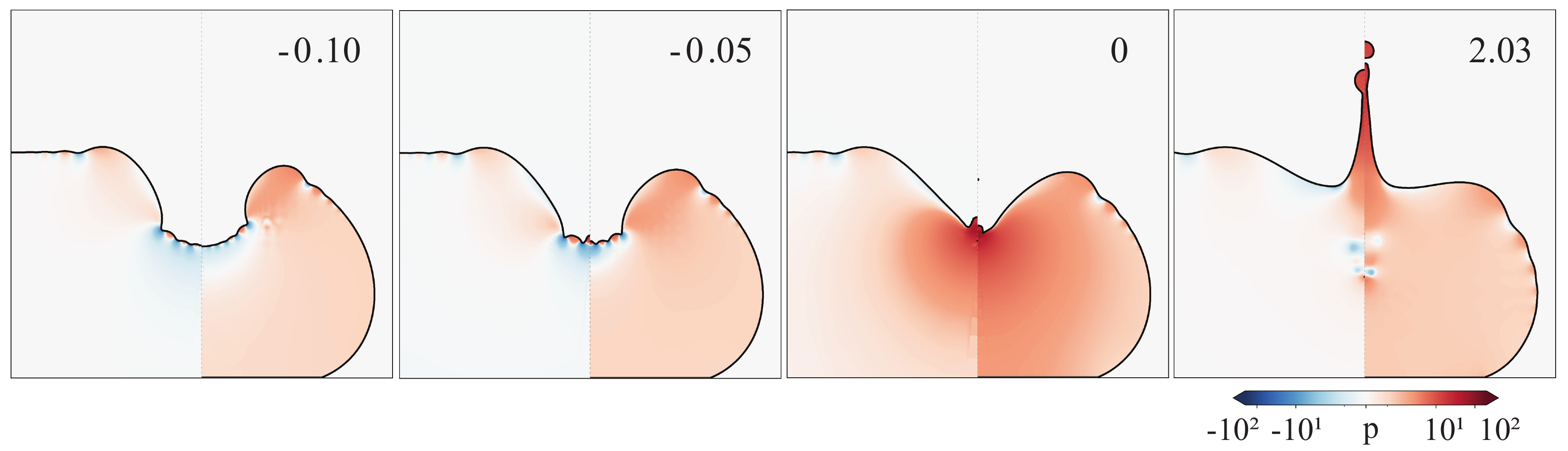}}
\caption{Hydrostatic pressure distribution during the cavity collapse in an infinite bath (left) and in a sessile droplet with $H/R_b=3.94$ and $R_D/R_b=1.99$ (right). The label indicates the time to the free surface reversal. The simulation was conducted for $\text{Bo}=0.143$ and  $\text{La}=6.9\times10^{4}$.}
\label{Fig_s1}
\end{figure*}

\begin{figure}
\resizebox{0.5\textwidth}{!}{\includegraphics{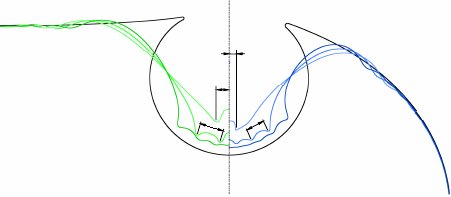}}
\caption{Cavity collapse in an infinite bath (green lines) and in a sessile droplet with $H/R_b=3.94$ and $R_D/R_b=1.99$ (blue lines). The black line is the initial bubble shape, and the green and blue lines correspond to the time to reversal equal to -0.10, -0.05, and 0. The arrows indicate the wavelengths and the size of the cavity bottom. The simulation was conducted for $\text{Bo}=0.143$ and $\text{La}=6.94\times10^{4}$.}
\label{Fig_s20}
\end{figure}

\begin{figure*}
\resizebox{0.8\textwidth}{!}{\includegraphics{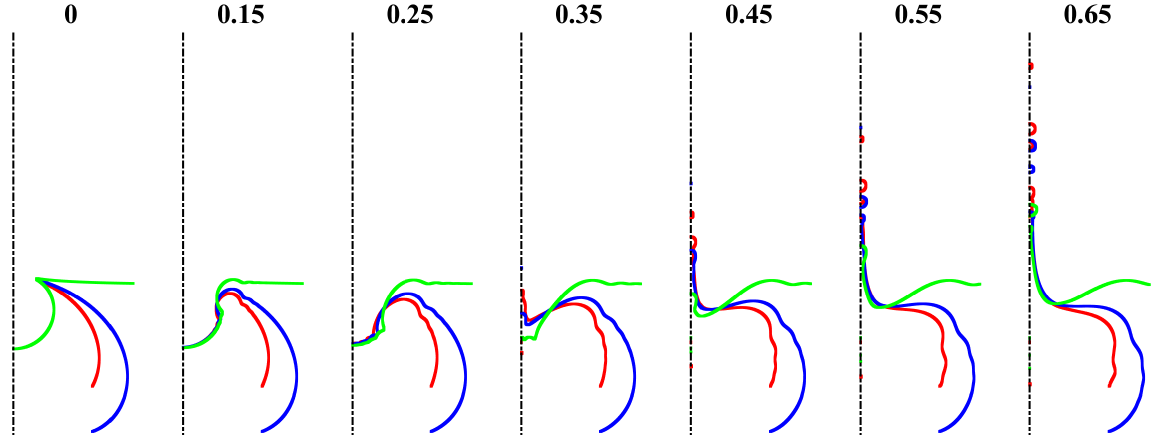}}
\caption{Cavity collapse in an infinite bath (green lines) and in a sessile droplet with $R_D/R_b=1.99$ and $H/R_b=2.82$ (red lines) and $3.94$ (blue lines). The label indicates the time from the start of the simulation. The simulation was conducted for $\text{Bo}=0.143$ and  $\text{La}=6.94\times10^{4}$.}
\label{Fig_s2}
\end{figure*}

\begin{figure}
\resizebox{0.37\textwidth}{!}{\includegraphics{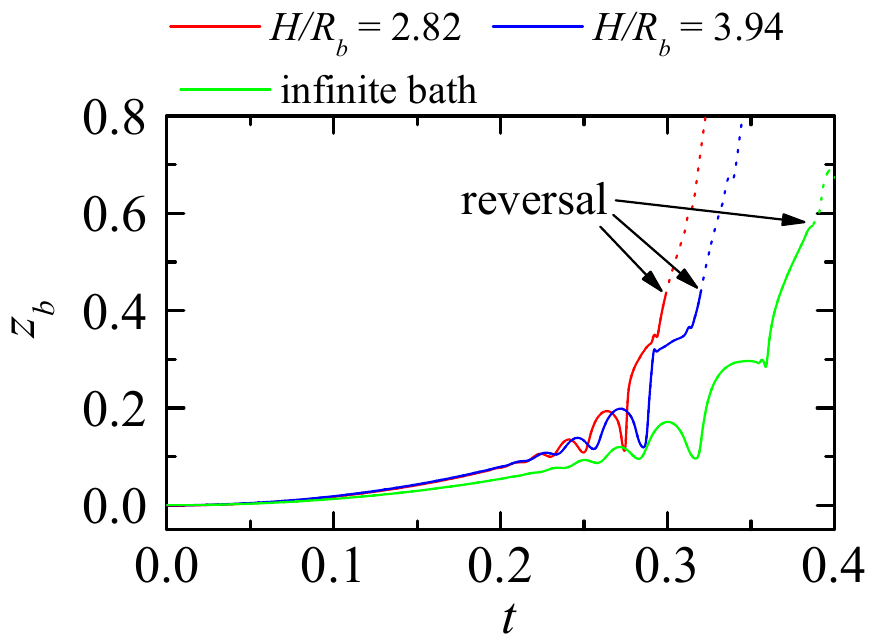}}
\caption{Cavity bottom height $z_b$ during cavity collapse in an infinite bath (green lines) and in a sessile droplet with $R_D/R_b=1.99$ and $H/R_b=2.82$ (red lines) and $3.94$ (blue lines). The simulation was conducted for $\text{Bo}=0.143$ and  $\text{La}=6.94\times10^{4}$. The first point on the dashed line indicates the instant after which the cavity bottom no longer shrinks, which can be approximated as the free surface reversal leading to jet formation.}
\label{Fig_Shape_t0}
\end{figure}

\begin{figure*}
\resizebox{0.7\textwidth}{!}{\includegraphics{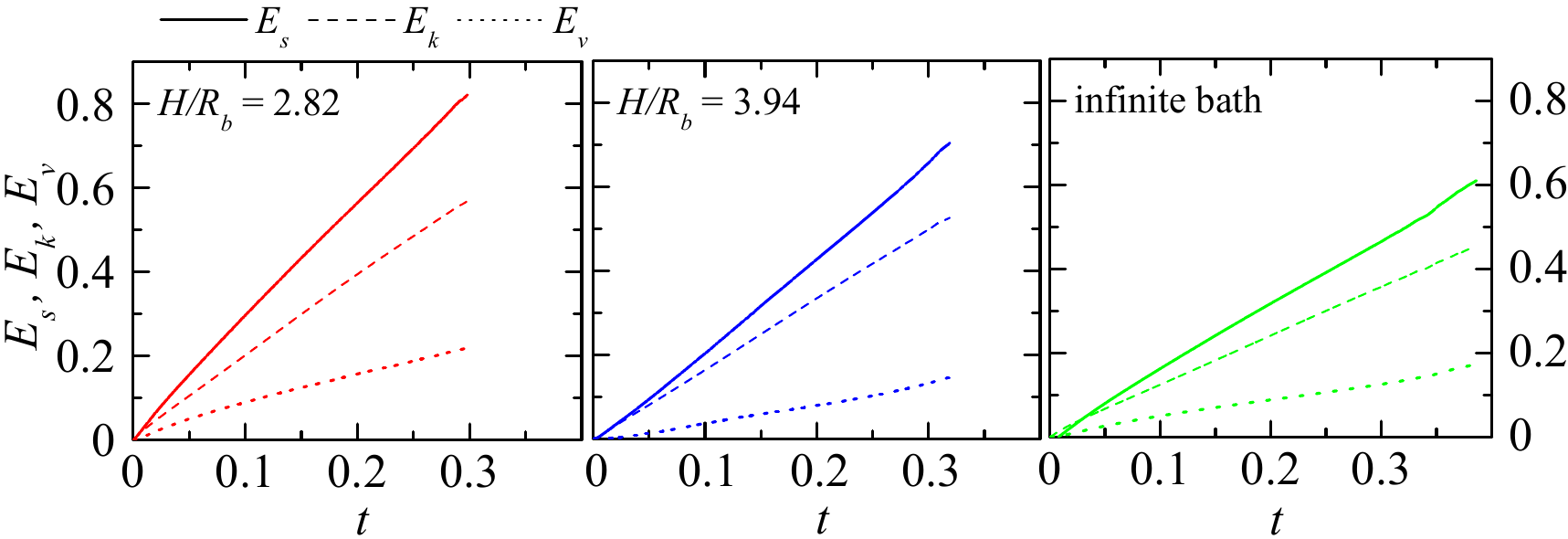}}
\caption{Released interfacial energy $E_s$, flow kinetic energy $E_k$, and energy dissipated by viscosity $E_v$ in a sessile droplet with $R_D/R_b=1.99$ and an infinite bath. The simulation was conducted for $\text{Bo}=0.143$ and  $\text{La}=6.94\times10^{4}$.}
\label{Fig_Shape_t10}
\end{figure*}

Figures \ref{Fig_s1}-\ref{Fig_Shape_t10} compare the bubble bursting in a sessile droplet with that in an infinite bath for the same values of the Bond and Laplace numbers. As observed in the experiments, the sessile droplet curvature increases the hydrostatic pressure by an amount of at least the order of the bubble capillary pressure $\sigma/R_b$ ($1\lesssim p\lesssim 10$ for $t=-0.10$). As a consequence, there is a significant pressure gradient that drives the liquid towards the bottom of the collapsing cavity (Fig.\ \ref{Fig_s1}). This effect compresses the bottom of the cavity (Fig.\ \ref{Fig_s20}), reducing the size of the emitted jet.  

In the sessile droplet case, capillary waves with smaller wavelengths and amplitudes arrive at the cavity bottom before jet formation (Fig.\ \ref{Fig_s20}). Consequently, smaller regions of low and high hydrostatic pressure arise near the cavity bottom in this case (Fig.\ \ref{Fig_s1}). This behavior is expected to favor energy focusing during jet formation, thereby reducing the jet's size and increasing its velocity. This effect is similar to that observed in a shallow liquid layer, where the proximity of a solid surface enhances capillary-wave damping near the cavity bottom \citep{YSCF26}.

Figure \ref{Fig_s2} compares the evolution of the interface shape during the cavity collapse in an infinite bath and in a sessile droplet. Interestingly, the cavity bottom in the sessile droplet is displaced upward relative to its position in the infinite bath at the same instant. This displacement increases as confinement increases ($H/R_b$ decreases). However, the jet forms at a lower height in the sessile droplet. This occurs because the jet forms faster in this case. This can be observed in Fig.\ \ref{Fig_Shape_t0}, which compares the cavity bottom height $z_b$ during the cavity collapse in an infinite bath and in a sessile droplet. The first points on the dashed lines correspond to the instants at which the cavity bottom no longer shrinks, which can be approximated as the beginning of jet formation. As mentioned above, $z_b$ takes larger values in the sessile droplet cases. However, the jet forms at a greater height in the infinite bath. 

Figure \ref{Fig_Shape_t10} shows the interfacial energy $E_s(t)=\sigma(\Sigma(0)-\Sigma(t))$ released by the gas-liquid interface $\Sigma$. This energy is distributed into the flow kinetic energy, $E_k(t)=\int_{\Omega} 1/2\, \rho\, v(t)^2\, d\Omega$, and the energy dissipated by viscosity, $E_v(t)=\int_0^t \int_{\Omega} {\bf \tau}(t'):{\bf \nabla}{\bf v}(t')\, d\Omega\, dt'$. Here, $\Omega$ is the fluid domain, ${\bf \tau}$ is the viscous stress tensor, and ${\bf v}$ is the velocity field. Both the released interfacial energy $E_s$ and the dissipated energy $E_v$ increase as $H/R_b$ decreases (i.e., as confinement increases). $E_k$ slightly increases with confinement despite the fraction of $E_s$ converted into $E_k$ decreases. This slight increase in $E_k$ does not explain the increase in droplet velocity, but rather reflects an enhancement in energy focusing.

\begin{figure*}
\resizebox{0.8\textwidth}{!}{\includegraphics{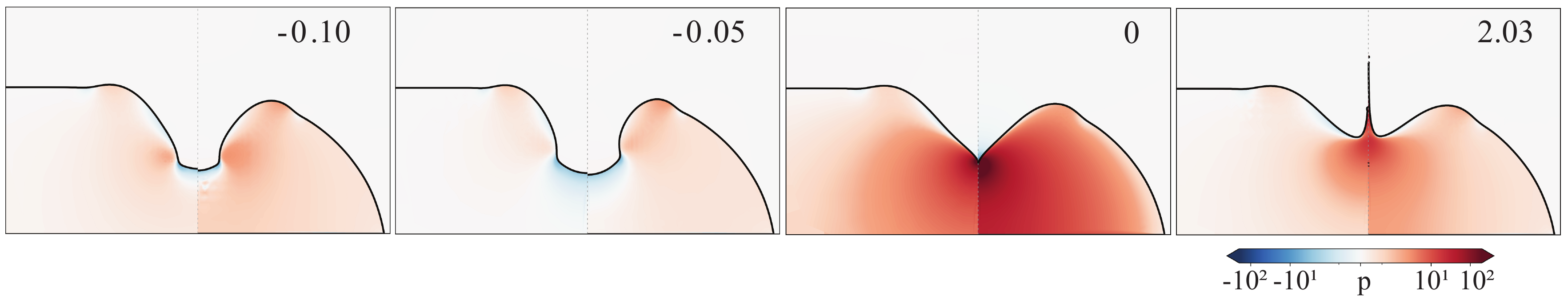}}
\caption{Hydrostatic pressure distribution during cavity collapse in an infinite bath (left) and in a sessile droplet with $R_D/R_b=2.57$ and $H/R_b=2.96$ (right). The labels indicate the time to the free surface reversal. The simulation was conducted for $\text{Bo}=0.048$ and  $\text{La}=468$.}
\label{Fig_s1v}
\end{figure*}

\begin{figure*}
\resizebox{0.8\textwidth}{!}{\includegraphics{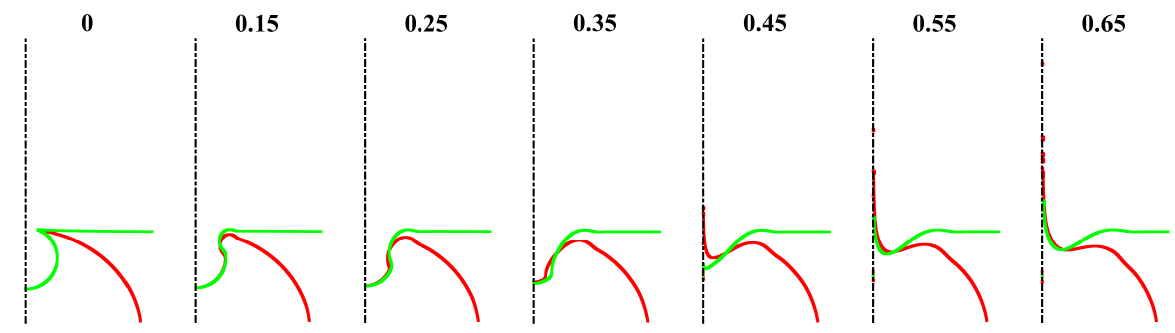}}
\caption{Comparison between the cavity collapse in an infinite bath (green lines) and in a sessile droplet with $R_D/R_b=2.57$ and $H/R_b=2.96$ (red lines). The label indicates the time to the free surface reversal. The simulation was conducted for $\text{Bo}=0.048$ and  $\text{La}=468$.}
\label{Fig_ShapeVisc}
\end{figure*}

We finish our numerical analysis by considering a viscous case. The cavity dynamics in the viscous (low Laplace number) regime substantially differ from those described above. Viscous stresses suppress the capillary waves with shorter wavelengths, improving the energy of transfer from the driving wave to the emitted jet. This enhanced focusing effect leads to thinner and faster jets than those produced in the low-viscosity case. This behavior is also observed in the sessile droplet (Fig.\ \ref{Fig_s1v}). As in the low-viscosity case, the droplet interface curvature produces an additional pressure gradient that drives the liquid towards the bottom of the bubble (Fig.\ \ref{Fig_s1v}) and reduces the time elapsed between cap film breakage and free-surface reversal (Fig.\ \ref{Fig_ShapeVisc}). Contrary to the low-viscosity case, this does not significantly shift the bottom upward. As observed in the experiments, energy focusing is enhanced in the sessile droplet, thereby enabling ejection below the Laplace number threshold in the infinite bath. In fact, Fig.\ \ref{Fig_ShapeVisc} shows that the sessile droplet emits a droplet for values of Bo and La at which ejection does not occur in an infinite bath.

\begin{figure}
\resizebox{0.35\textwidth}{!}{\includegraphics{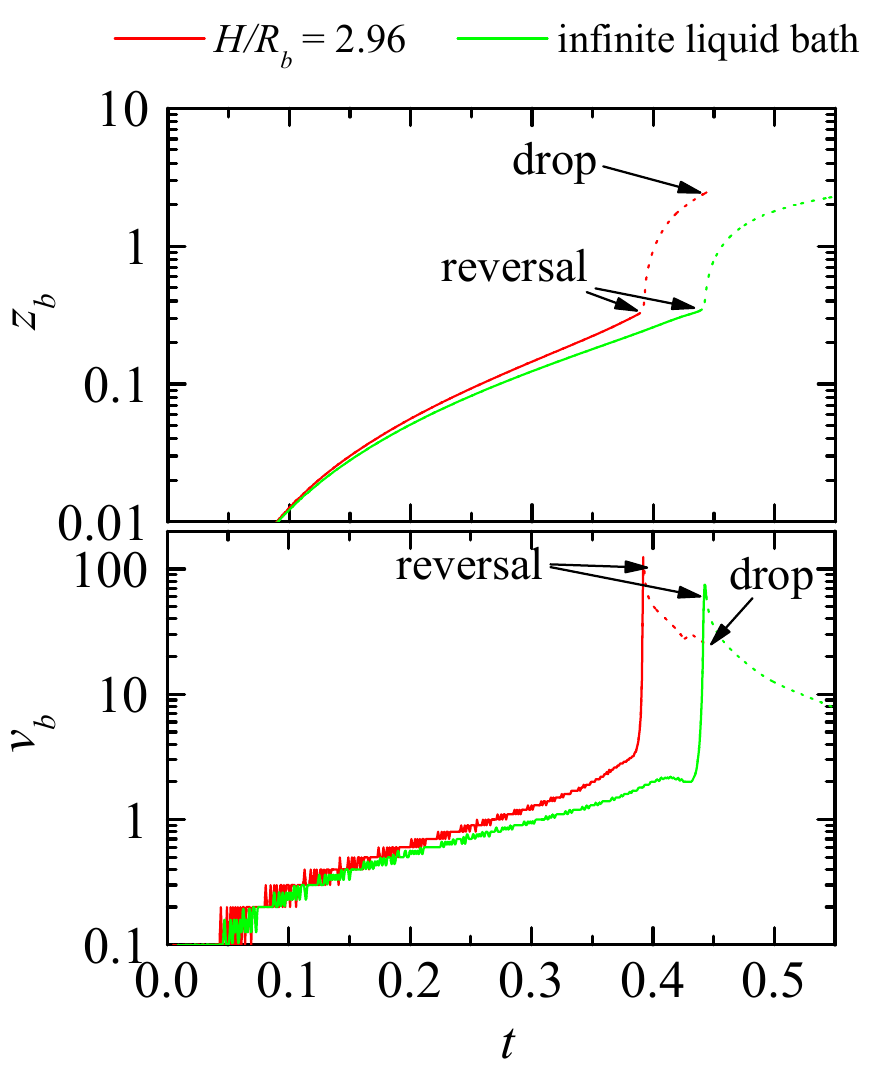}}
\caption{Vertical position $z_b$ and velocity $v_b$ of the cavity bottom in an infinite bath (green lines) and in a sessile droplet with $R_D/R_b=2.57$ and $H/R_b=2.96$ (red lines). The first point on the dashed line indicates the instant after which the cavity bottom no longer shrinks, which can be approximated as the free surface reversal leading to jet formation. The simulation was conducted for $\text{Bo}=0.048$ and  $\text{La}=468$.}
\label{Fig_ShapeViscXX}
\end{figure}

Figure \ref{Fig_ShapeViscXX} shows the vertical position $z_b$ and velocity $v_b$ of the cavity bottom in an infinite bath and in a sessile droplet. In this latter case, the last point of the curve corresponds to the droplet ejection. The bottom sharply accelerates at the free surface reversal in both configurations, leading to jet formation. The velocity reached in the sessile droplet case is around 20\% larger than in the infinite bath case. This increase is sufficient to produce the droplet detachment. 

\section{Concluding remarks}
\label{sec7}

After studying the bursting of a bubble within a sessile droplet, we experimentally found that the bubble can emit jet droplets for Laplace numbers below $\approx 300$, a value significantly smaller than the minimum one, $\text{La}\simeq 540$, in an infinite liquid bath. This implies that bubbles with radii smaller than $\approx 4$ $\mu$m can produce water jets from sessile droplets (the minimum radius for an infinite bath is $\approx 7.5$ $\mu$m). This is probably the major result of the present work. Moreover, our numerical results for $\text{La}\simeq 468$ show that the droplet interface curvature produces an additional pressure gradient that drives the liquid towards the bottom of the bubble, which reduces the time elapsed between cap film breakage and free-surface reversal. As observed in the experiments, this enhances energy focusing and enables the ejection of a droplet below the Laplace number threshold for an infinite bath.

In the high-Laplace number regime, confinement reduces the length of capillary waves that travel towards the bubble bottom before the free-surface reversal. Notably, the amplitude of these waves decreases in the sessile-droplet case, a trend linked to improved energy focusing. As in the viscous case, the pressure gradient arising from droplet curvature drives the liquid towards the bottom of the bubble. This shifts the bottom upward and reduces the time elapsed between cap film breakage and free-surface reversal. These effects result in the ejection of thinner, faster jets than in the infinite bath case.

Our study identifies geometric confinement and free surface curvature as key factors controlling aerosol production due to bubble bursting. Specifically, we have described a mechanism that allows the ejection of jet droplets for experimental conditions where an infinite bath cannot emit them. Our results may be relevant to a wide range of practical situations, including the fluid dynamics of disease transmission and industrial applications such as printing, drug delivery and bioprinting. The phenomenon revealed here provides new insights into aerosol generation and may offer a basis for the rational design of surfaces aimed at controlling aerosol emission from bursting bubbles.

\vspace{1cm}

The data that support the findings of this study are available from the corresponding author upon reasonable request. 

\section*{Acknowledgement}
This work has been 85\% co-financed by the European Union, the European Regional Development Fund, and the la Junta de Extremadura. Autoridad de Gestión. Ministerio de Hacienda (GR24077). This publication is part of grant FJC2021-046447-I, financed by MCIN/AEI/10.13039/501100011033 and European Union ‘‘NextGenerationEU’’/PRTR. U.J.G.-H. acknowledges support from DGAPA through Subprograma de Incorporación de Jóvenes Académicos de Carrera (SIJA) and the grant PAPIIT-UNAM IA103726 and IN114325. DFR acknowledges funding from the project Needle-free injections with file number 19657 of the research programme NWO Talent Programme Vidi TTW, which is financed by the Dutch Research Council (NWO).


%

\end{document}